%% file: template.tex
\documentclass{Interspeech}

\interspeechcameraready

\title{Pronunciation Editing for Finnish Speech  using Phonetic Posteriorgrams}

\author{Zirui}{Li}
\author{Lauri}{Juvela}
\author{Mikko}{Kurimo}

\affiliation[nocounter]{Department of Information and Communication Engineering}{Aalto University}{Finland}
\email{zirui.li@aalto.fi, lauri.juvela@aalto.fi, mikko.kurimo@aalto.fi} 

\usepackage{comment}

\begin{document}

\maketitle

\begin{abstract}
    
    
    Synthesizing second-language (L2) speech is potentially highly valued for L2 language learning experience and feedback. However, due to the lack of L2 speech synthesis datasets, it is difficult to synthesize L2 speech for low-resourced languages. 
    In this paper, we provide a practical solution for editing native speech to approximate L2 speech and present PPG2Speech, a diffusion-based multispeaker Phonetic-Posteriorgrams-to-Speech model that is capable of editing a single phoneme without text alignment. We use Matcha-TTS's flow-matching decoder as the backbone, transforming Phonetic Posteriorgrams (PPGs) to mel-spectrograms conditioned on external speaker embeddings and pitch. PPG2Speech strengthens the Matcha-TTS's flow-matching decoder with Classifier-free Guidance (CFG) and Sway Sampling. We also propose a new task-specific objective evaluation metric, the Phonetic Aligned Consistency (PAC), between the edited PPGs and the PPGs extracted from the synthetic speech for editing effects. 
    We validate the effectiveness of our method on Finnish, a low-resourced, nearly phonetic language, using approximately 60 hours of data. We conduct objective and subjective evaluations of our approach to compare its naturalness, speaker similarity, and editing effectiveness with TTS-based editing. Our source code is published at \url{https://github.com/aalto-speech/PPG2Speech}.
    
\end{abstract}

\input{sections/introduction}

\input{sections/Background}

\input{sections/Method}

\input{sections/Experiments}

\input{sections/Conclusion}

\input{sections/acknowledge}

\newpage

\bibliographystyle{IEEEtran}
\bibliography{mybib}

\end{document}

%% file: sections/introduction.tex
\section{Introduction}

Proficiency in second languages (L2) enhances personal communication skills and cultural understanding, and it is also an essential criterion for evaluating immigrants' integration into the local society. Pronunciation, as a challenging yet essential part of language learning, benefits from pronunciation feedback from a teacher~\cite{Derwing2015}. Recent studies demonstrate that speech synthesis technology effectively supports L2 pronunciation training by providing individual pronunciation feedback~\cite{Liakin19052017}, and providing synthesized L2 pronunciation to the learner improves learning experience and results~\cite{cardoso2018learning}.

L2 speech synthesis can be treated as a subdomain of accented speech synthesis. Although current research in accented speech synthesis~\cite{10439064} works well with limited data, it still requires the recording of accented speech. Therefore, synthesizing L2 speech is challenging for low-resourced languages, like Finnish, due to the scarcity of suitable L2 speech synthesis data. This scarcity necessitates exploring methods such as phoneme-level speech editing, where native speech is systematically modified to approximate specific L2 pronunciations.


Speech editing is a technique that allows insertion, deletion, and replacement of certain speech content. Current speech editing models~\cite{tan2021editspeech,wang2022campnet,jiang2023fluentspeech, liu2024fluenteditor} adopt a mask-and-infill approach, which consists of masking the edited region, filling this region with new content, and ensuring prosody consistency by imposing auxiliary loss on the edge of the edited region. These methods are text-based, operate at the word-level, and depend on external aligners to provide alignment information to determine the edited region. Early speech editing methods~\cite{tan2021editspeech,wang2022campnet} are trained with reconstruction loss on the mel-spectrogram, making it less ideal in synthesis quality. Recent editing methods~\cite{jiang2023fluentspeech, liu2024fluenteditor} are diffusion-based, improving the synthesized sample quality.

Diffusion-based Speech Synthesis using Flow Matching~\cite{lipman2023flow} has become a popular research topic in recent years~\cite{le2023voicebox, ju2024naturalspeech, matchatts}. As a class of Non-Autoregressive (NAR) models, flow-matching-based Speech Synthesis models benefit from parallel computation of all temporal steps and can effectively balance the synthesis quality and the inference speed by adjusting the number of diffusion sampling steps. However, NAR models require modeling the alignment between input text and output speech. Voicebox~\cite{le2023voicebox} and NaturalSpeech 3~\cite{ju2024naturalspeech} rely on an external phoneme-level aligner that expands the input phoneme sequence to the length of the synthesized speech based on the duration of each phoneme, before inputting the phoneme sequence to the model. Matcha-TTS~\cite{matchatts} uses monotonic alignment search to find the alignment between phoneme representations and synthesized speech, and trains a duration predictor to predict the duration of each phoneme. All these methods~\cite{le2023voicebox, ju2024naturalspeech, matchatts} impose hard alignments between input text and synthesized speech, which might hurt the naturalness of the synthesized speech~\cite{eskimez2024e2}.


To gain more control over the pronunciation information, we use Phonetic Posteriorgrams (PPGs) as the input, instead of the phoneme sequence. PPGs are time-varying categorical distributions over phonemes~\cite{ppgs}. They have been widely used for different speech processing tasks~\cite{6639269, 5372889, 7552917, miyoshi2017voice, 9414137, morrison2024fine} in the past years. As a time-varying distribution over phoneme categories, they contain rich pronunciation information that makes them ideal for pronunciation assessment and mispronunciation detection~\cite{6639269}, and spoken term detection~\cite{5372889}. PPGs also disentangle well with the timbre and excitation of speech, making them ideal as content representations for speech and singing voice conversion~\cite{7552917, miyoshi2017voice, 9414137}. Furthermore, they are interpretable representations for humans and can be potentially used for neural speech editing~\cite{morrison2024fine}.

In this paper, we present PPG2Speech, a diffusion-based multispeaker PPG-to-speech model using Matcha-TTS's flow-matching decoder~\cite{matchatts}, enhanced with classifier-free guidance (CFG)~\cite{ho2021classifierfree} and sway sampling~\cite{chen-etal-2024-f5tts}. Our proposed approach operates without textual input, providing a practical solution for editing native speech to approximate common mispronunciations made by L2 learners. Unlike the previous approaches in Speech Editing and TTS, which require explicit hard alignment for model training and determining the edited region, our model uses PPGs, which provide soft, explicit alignment information that retains the naturalness of synthetic speech. We validate our method with both objective and subjective evaluations to assess naturalness, speaker similarity, and editing effectiveness in comparison with TTS-based editing techniques. Objective metrics include a novel Phonetic Aligned Consistency (PAC) to assess editing effectiveness.

%% file: sections/Background.tex
\section{Background}



\subsection{Flow matching for generative models}

Flow Matching (FM) is a recent generative modeling technique designed to transform a simple prior distribution, such as a standard Gaussian, into a complex data distribution through deterministic flows defined by ordinary differential equations (ODEs)~\cite{lipman2023flow}. Given an unknown data distribution \( q(x) \), FM constructs a probability density path \( p_t(x) \), with \( t \in [0,1] \), smoothly transitioning from a known prior distribution \( p_1(x) = \mathcal{N}(x;0,I) \), a standard Gaussion distribution, to a final distribution \( p_0(x) \) approximating the data \( q(x) \).

Specifically, FM defines a neural-parameterized vector field \( v_t(x; \theta) \) that generates a flow \( \phi_t(x) \) through the ODE:
\begin{equation}
\label{eqn:flow_matching_ode}
\frac{d}{dt}\phi_t(x) = v_t(\phi_t(x); \theta), \quad \phi_0(x) = x.
\end{equation}
The FM objective minimizes the difference between this parameterized vector field \( v_t(x; \theta) \) and a known conditional vector field \( u_t(x|x_1) \) that generates a conditional probability path \( p_t(x|x_1) \) via a regression loss:
\begin{equation}
\label{eqn:cfm_loss}
L_{\text{CFM}}(\theta) = \mathbb{E}_{t, q(x_1), p_t(x|x_1)}\left[ \\|| v_t(x; \theta) - u_t(x|x_1) \\||^2 \right].
\end{equation}

Optimal Transport Conditional Flow Matching (OT-CFM) simplifies this further by defining the flow explicitly as linear interpolation between noise samples \( x_1 \sim p_1(x) \) and data samples \( x_0 \sim q(x) \), such that:
\begin{equation}
\label{eqn:ot_linear_interpolation}
\phi_t^{\text{OT}}(x) = t x_1 + (1 - t) x_0,
\end{equation}
leading to the simple and computationally efficient OT-CFM loss:
\begin{equation}
\label{eqn:ot_cfm_loss}
    L_{\text{OT-CFM}}(\theta) = \mathbb{E}_{t,q(x_1),p_0(x_0)}\left[\\||v_t(\phi_t^{\text{OT}}(x);\theta) - (x_0 - x_1)\\||^2\right].
\end{equation}
where the gradient vector field $u_t(x|x_1) = (x_0 - x_1)$ is linear, time-invariant, and only depends on $x_0$ and $x_1$. This leads to faster training, sampling, and better generalization compared to traditional Diffusion Probabilistic Models (DPMs).

\subsubsection{Classifier-free guidance}

Classifier-free guidance is a technique developed to enhance sample fidelity in conditional diffusion models without relying on an external classifier~\cite{ho2021classifierfree}. The key idea is to jointly train a conditional and an unconditional diffusion model using a shared neural network. During training, the model is conditioned on the input conditions $c$ with some probability $P_\text{con}$, and with a complementary probability $P_\text{uncon} = 1 - P_\text{con}$, it is trained unconditionally by providing a null condition. This allows the network to learn both conditional and unconditional score estimates $v_t(\phi_t(x_t), c)$ and $v_t(\phi_t(x_t))$, respectively.

At inference time, the two score estimates are linearly combined to form a modified guided score:
\begin{equation}
\label{eqn:cfg}
    v_{t,\text{CFG}} = v_t(\phi_t(x_t), c) + w \left[v_t(\phi_t(x_t), c) - v_t(\phi_t(x_t))\right].
\end{equation}
where $w \geq 0$ is a guidance strength hyperparameter that controls the trade-off between sample quality and diversity. A higher $w$ increases fidelity at the cost of diversity.

\subsubsection{Sway Sampling}

Typically, $n$ timesteps are sampled uniformly for $t \in [0, 1]$ during the diffusion sampling process, where $n$ is the total number of iterations in the process. Recent research~\cite{chen-etal-2024-f5tts} demonstrates the use of Sway Sampling improves the quality of the sample. The Sway Sampling is formulated as:
\begin{equation}
\label{eqn:sway_samp}
f_{\text {sway }}(u ; s)=u+s \left(\cos \left(\frac{\pi}{2} u\right)-1+u\right),
\end{equation}
where the sway coefficient $s \in\left[-1, \frac{2}{\pi-2}\right]$ and $u \sim \mathcal{U}[0, 1]$. When $s < 0$, it takes smaller sampling steps for the early stage sampling and larger steps for the later stages. When $s > 0$, it takes larger sampling steps for the early stages and smaller steps for the later stages. It becomes uniform sampling when $s = 0$.

%% file: sections/Method.tex
\section{Method}

\subsection{PPG2Speech}
\begin{figure}[t]
  \centering
  \includegraphics[width=\linewidth]{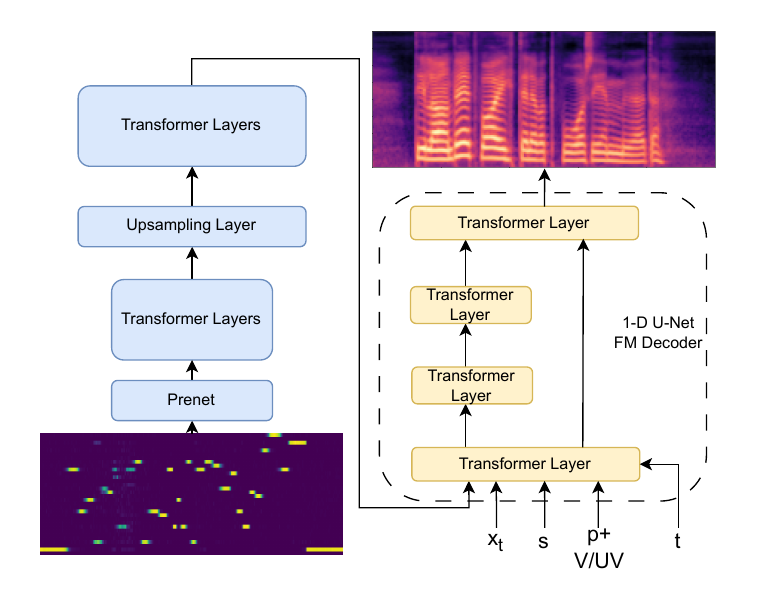}
  \caption{The diagram of our model. $x_t$ is the noisy mel-spectrogram. $s$ is the speaker embedding, $p+V/UV$ is the pitch embedding sequence concatenated with the voiced/unvoiced flag, and $t \in [0, 1]$ is the diffusion time step.}
  \label{fig:model}
\end{figure}

We show our PPG2Speech model's architecture in Figure~\ref{fig:model}. Our model generates mel-spectrograms from PPGs with conditional pitch sequences and speaker embeddings. The model consists of a PPG encoder and a flow-matching decoder. The PPG encoder consists of a prenet, a stack of Transformer/Conformer~\cite{gulati2020conformer} layers, with an upsampling layer in between. The prenet maps the PPG channels to the Transformer's channel dimensions. The upsampling layer upsamples the latent PPG representations to the time resolution of the target mel-spectrogram based on the nearest neighbor method. The encoder uses Relative Positional Encoding~\cite{dai2019transformer} to provide positional information to the self-attention.

For the flow-matching decoder, we extend the 1-D U-Net based decoder in Matcha-TTS~\cite{matchatts} with Classifier-free Guidance (CFG). Furthermore, the decoder is conditioned on external speaker embeddings and pitch embedding sequences to generalize to unseen speakers and synthesize more natural speech. The original speaker embedding table in Matcha-TTS is discarded. During training, we uniformly sample the timestep $t \in [0, 1]$ to predict the gradient vector field. During inference, we use Sway Sampling with $s = -1$ to transform the speech from the noise using smaller steps in the initial steps to improve the quality of the synthesized speech. The synthesized mel-spectrogram is converted to waveform using a pretrained HiFi-GAN vocoder~\cite{kong2020hifi}.

\subsection{Phonetic Aligned Consistency}
\label{sec:pronun}

Given the synthetic speech from the edited PPG or edited text, we first extract the PPG from it and then find the region that corresponds to the phoneme-level editing. Then we calculate the Phonetic Aligned Consistency (PAC) as:

\begin{equation}
\label{eqn:dtw_score}
\text{PAC}_{1:m,1:n} = 
\frac{1}{m} 
 \text{DTW}_{\text{JSD}}(\text{PPG}^{m}_{\text{edited}}, \text{PPG}^{n}_{\text{syn}}),
\end{equation}
where $\text{PPG}^{m}_{\text{edited}}$ is the edited region of the edited PPG with length $m$, $\text{PPG}^{n}_{\text{syn}}$ is the edited region in the PPG of the synthetic speech with length $n$, and $\text{JSD}$ is the Jensen-Shannon Distance. We normalize the score by the length of the edited region in the edited PPG.

Our motivation for Equation~\ref{eqn:dtw_score} are:

\begin{enumerate}
    \item PPG is a time-varying distribution over phoneme categories. Therefore, it is natural to use the JSD to measure the discrepancy between the corresponding frames in two PPGs. Other probability distance metrics are also viable in Equation~\ref{eqn:dtw_score}.
    \item We use DTW to tackle the length differences between the $\text{PPG}^{m}_{\text{edited}}$ and the $\text{PPG}^{n}_{\text{syn}}$.
    \item DTW will naturally cause a larger cost for longer edited timespan, hence we normalize the cost by the length of $\text{PPG}^{m}_{\text{edited}}$ to compensate for the larger costs from longer edits, so that the scores are comparable for different edits on different utterances.
\end{enumerate}

%% file: sections/Experiments.tex
\section{Experiments}

\subsection{Data}
\label{sec:data}

We conduct our experiments on the combination of two Finnish datasets, \textbf{Perso Synteesi}~\cite{perso} and \textbf{Finsyn}~\cite{finsyn}. We show the details of these two datasets in Table~\ref{tab:data_spec}. \textbf{Perso Synteesi}~\cite{perso} is a read-speech dataset collected from 33 male native speakers and 33 female native speakers by recording speakers' voices when reading prompts. Prompts are the same for all speakers. We only have the data of 20 male speakers and 30 female speakers. \textbf{Finsyn}~\cite{finsyn} is another Finnish speech synthesis dataset with two female speakers and various speaking styles. We follow the suggestion from~\cite{finsyn} and filter out those utterances with a spontaneous speaker style to ensure stable synthesis results.

\begin{table}[htbp]
\centering
\caption{Dataset details for \textbf{Perso Synteesi} and \textbf{Finsyn}.}
\begin{tabular}{|c|ccc|}
\hline
\textbf{Dataset}                         & \textbf{Gender} & \textbf{\# of speakers} & \textbf{Hours} \\ \hline
\multirow{3}{*}{\textbf{Perso Synteesi}} & Male            & 20 & 6.8            \\
                                         & Female          & 30 & 10.7           \\
                                         & Total           & 50 & 17.6           \\ \hline
\multirow{3}{*}{\textbf{Finsyn}}         & Male            & 0 & 0              \\
                                         & Female          & 2 & 44.3           \\
                                         & Total           & 2 & 44.3           \\ \hline
\textbf{Total} & - & 52 & 61.8 \\ \hline
\end{tabular}
\label{tab:data_spec}
\end{table}

We split our data into \textbf{Train}, \textbf{Validation}, \textbf{Test}, and \textbf{Test\_unseen} sets. For the \textbf{Perso Synteesi}~\cite{perso}, we use the original partition of the dataset, and exclude speakers $19$, $20$, $19m$, $20m$ from the training and validation set and use their original validation and test set as \textbf{Test\_unseen} for testing the model's generalization on unseen speakers. For the \textbf{Finsyn}~\cite{finsyn} dataset, we split the data into training, validation, and testing sets with ratio $0.9:0.05:0.05$. No utterance from \textbf{Finsyn} is split to \textbf{Test\_unseen} because \textbf{Finsyn} only has 2 speakers. The resulting splits are shown in Table~\ref{tab:data_split}.

We extract the PPGs from the Kaldi HMM-TDNN-medium model reported in~\cite{kaldi_model}, which contains 29 Finnish phonemes, and 3 special symbols, \textit{SPN}, \textit{SIL}, and \textit{eps}. We use the pretrained \textit{SimAMResNet34}\footnote{\url{https://github.com/wenet-e2e/wespeaker/blob/master/docs/pretrained.md}} model from WeSpeaker toolkit~\cite{wang2023wespeaker} to extract the speaker embeddings, and use the pretrained Pitch Estimation Neural Network (PENN)~\cite{morrison2023cross} to extract pitch and periodicity from the audio. We standardize the log pitch per utterance by subtracting its mean and dividing its variance to mitigate timbre leakage, and then quantize pitch into 256 bins. The mel-spectrograms are extracted from the audio using HiFi-GAN~\cite{kong2020hifi} configurations.

\begin{table}[htbp]
\centering
\caption{Hours and the number of speakers in different splits of the dataset.}
\begin{tabular}{|l|ll|}
\hline
\textbf{Splits}       & \textbf{\# of speakers} & \textbf{Hours} \\ \hline
\textbf{Train}        &    48       & 53.6           \\
\textbf{Validation}   &    48       & 3.9            \\
\textbf{Test}         &    48       & 2.7            \\
\textbf{Test\_unseen} &    4        & 0.15           \\ \hline
\end{tabular}
\label{tab:data_split}
\end{table}

\begin{table*}[t!]
\centering
\caption{The objective results. \textbf{SECS} is the Speaker Encoder Cosine Similarity, the Character Error Rate (\textbf{CER}) is used to evaluate intelligibility. Furthermore, Mel-Cepstral Distortion (\textbf{MCD}) and pitch mean absolute error (\textbf{Pitch MAE}) are used to evaluate the model's capability on unseen speakers. The Phonetic Aligned Consistency (\textbf{PAC}) is the objective evaluation for editing effects.}
\begin{tabular}{c|ccccccc}
\textbf{Model}             & \textbf{SECS}$\uparrow$ & \textbf{CER}$\downarrow$ & \textbf{\begin{tabular}[c]{@{}c@{}}SECS\\ Unseen\end{tabular}}$\uparrow$ & \textbf{\begin{tabular}[c]{@{}c@{}}CER\\ Unseen\end{tabular}}$\downarrow$ & \textbf{\begin{tabular}[c]{@{}c@{}}MCD\\ Unseen\end{tabular}}$\downarrow$ & \textbf{\begin{tabular}[c]{@{}c@{}}Pitch MAE\\ Unseen $\Delta\mbox{\textcent}$\end{tabular}}$\downarrow$ & \textbf{PAC}$\downarrow$ \\ \hline
\textbf{Groundtruth}       & 0.89          & 4.19         & 0.89                                                    & 3.36                                                  & -                                                             & -                                                                   & -                    \\ \hline
\textbf{Matcha-TTS}        & 0.87          & \textbf{3.74}         & 0.75                                                           & \textbf{3.80}                                                          & -                                                             & -                                                                   & -                    \\
\textbf{Matcha-TTS-CFG} & \textbf{0.89}          & 3.91         & 0.78                                                           & 3.91                                                          & -                                                             & -                                                                   &  0.804            \\ \hline
\textbf{PPG2Speech}             & 0.83          & 5.37         & 0.76                                                           & 4.00                                                          & 3.71                                                          & 9.49                                                                & -                    \\
\textbf{PPG2Speech-CFG}      & 0.86          & 4.70         & \textbf{0.81}                                                           & 3.82                                                          & \textbf{3.69}                                                          & \textbf{7.17}                                                                &  \textbf{0.709}      
\end{tabular}
\label{tab:obj_eval}
\end{table*}

\subsection{Model details}

We first use a 3-layer convolutional prenet with each layer having a kernel size 3 to map the 32-dim PPGs to 128-dim latent. We use the 2-layer Conformer~\cite{gulati2020conformer} before the upsampling layer with $4$ attention heads, 512 feed-forward dimension, and a convolution kernel size of 9. After the upsampling, we use a 2-layer Transformer with 4 attention heads and 512 feed-forward dimensions. We find this configuration yields the best outcome.

Our flow-matching decoder follows the same configuration as in~\cite{matchatts}. The external speaker embedding is of 256 dimensions. We use an embedding table of dimension 16 to represent the quantized pitch and concatenate the log periodicity with the pitch embedding as the soft voiced/unvoiced (V/UV) flag. External speaker embedding and pitch embedding with the V/UV flag are concatenated together as the conditional input to the decoder. During training, there is a $10\%$ probability for each batch that the PPG latent and the conditional input will be dropped to train the unconditioned model. During inference, we set the guidance strength $w = 3$, and the sway coefficient is set to $-1$.

We train both the Matcha-TTS and our PPG2Speech model from scratch using the training set and evaluate on the test set and the unseen test set. Both models are trained with a vanilla flow-matching decoder in Matcha-TTS and a flow-matching decoder strengthened by CFG, to show the CFG's effect on the model's performance. All models are trained with a batch size of 32 for 500k updates on 1 Nvidia A100 GPU. The maximum learning rate is 1e-4, and we use a cosine annealing learning rate scheduler with 150k steps linear warmup to adjust the learning rate during training.

\subsection{Evaluation \& Results}

\subsubsection{Objective Evaluation}

The results of the objective evaluation are shown in Table~\ref{tab:obj_eval}. We use Speaker Encoder Cosine Similarity (SECS) to measure the similarity between the groundtruth speaker embedding and the speaker embedding extracted from the synthesized speech using the same \textit{SimAMResNet34} model in Section~\ref{sec:data}. We use a Wav2Vec2-large model\footnote{\url{https://huggingface.co/GetmanY1/wav2vec2-large-fi-150k-finetuned}} that is pretrained on 158k hours of unlabeled Finnish speech and fine-tuned on 4600 hours of Finnish speech to evaluate the intelligibility of the synthesized speech and use Character Error Rate (CER) as the objective metric. Furthermore, we use Mel-Cepstral Distortion (MCD)~\cite{chen2022v2c} and pitch mean absolute error (Pitch MAE) to evaluate the model's capability of handling unseen speakers and how well the model follows the given pitch, respectively.


In Table~\ref{tab:obj_eval}, on the seen testset, the Groundtruth has an SECS of 0.89 and a CER of 4.19, while on the unseen testset it shows an SECS of 0.89 and a CER of 3.36. The vanilla Matcha-TTS achieves an SECS of 0.87 on seen data, and when enhanced with CFG, it improves to an SECS of 0.89, matching Groundtruth similarity. In comparison, the PPG2Speech method records an SECS of 0.83 and a CER of 5.37 on seen data; however, on unseen data, its performance drops to an SECS of 0.76, with an MCD of 3.71 and a Pitch MAE of 9.49\,\textcent. When CFG is applied to PPG2Speech, the unseen SECS increases to 0.86, the MCD decreases to 3.69, and the Pitch MAE decreases to 7.17\,\textcent. The comparison between models without CFG and models with CFG demonstrates the effectiveness of CFG on improving the quality of the synthesized speech. Furthermore, when comparing the baseline Matcha-TTS model and the PPG2Speech model, we see that Matcha-TTS outperforms the PPG2Speech model on the seen testset, yet PPG2Speech models perform better on unseen speaker similarity. Strengthening the PPG2Speech model with CFG achieves a comparable performance with the best CER (3.82 vs 3.80) results in the unseen testset. Our findings for objective evaluation can be summarised as follows:
\begin{enumerate}
    \item CFG improves the model's generalization to unseen speakers.
    \item Although we normalized and quantized pitch, timbre leak still exists and negatively affects the speaker similarity, as our PPG2Speech models have a lower SECS compared to their Matcha-TTS counterparts.
    \item PPG leads to higher CER, possibly due to its probability density nature and error propagation from the Kaldi TDNN model.
\end{enumerate}

\begin{table}[htbp]
\centering
\caption{Finnish phoneme editing table in our experiments. \textbf{Source} is the phoneme we selected from the PPGs/text, and \textbf{target} are the phonemes we edited to.}
\begin{tabular}{c|cccccc}
\textbf{Source} & ä    & ö & y    & r    & a & o \\ \hline
\textbf{Target} & a, e & o & u, e & l, w & ä & ö
\end{tabular}
\label{tab:edit}
\end{table}

To evaluate the editing effect objectively, we conduct the editing experiments on the Matcha-TTS with CFG model and the PPG2Speech model with CFG. We chose Matcha-TTS with CFG for this experiment because it is better than vanilla Matcha-TTS in terms of SECS with a small cost on CER, yet its CER on the seen testset is still better than the groundtruth.
Suggested by Finnish phoneticians, we edit the PPGs and their corresponding texts based on common pronunciation mistakes from L2 Finnish speakers, shown in Table~\ref{tab:edit}. The source phonemes in Table~\ref{tab:edit} are the correct phonemes in a word, and the target phonemes are mistakes that L2 speakers commonly make. We then calculate PAC between the edited PPGs and the PPGs estimated from the synthesized speech. The editing and evaluation steps are:
\begin{enumerate}
    \item Randomly select a source phoneme from original PPGs and replace it with the target phoneme by moving the probability density on the source phoneme to the target phoneme. If multiple target phonemes exist, we randomly select one for replacement.
    \item If the selected source phoneme is a part of a long vowel (has identical adjacent phonemes), edit the adjacent phonemes to the target phoneme as well.
    \item Edit the text by using the alignment information in the PPG to find the corresponding source phoneme. Then replace the source phoneme with the same target phoneme as in the PPG editing, and use the edited text as input for Matcha-TTS.
    \item Synthesize speech from edited PPG/edited text.
    \item Estimate the PPG using the same HMM-TDNN-medium model as Section~\ref{sec:data} and use Montreal-Forced-Aligner~\cite{mcauliffe17_interspeech} to find the corresponding edited region. Calculate the PAC in Section~\ref{sec:pronun} between the edited PPG and the estimated PPG in the edited region.
\end{enumerate}

\begin{table*}[th]
\centering
\caption{Table for subjective evaluation, besides Naturalness MOS (\textbf{NMOS}) and Speaker MOS (\textbf{SMOS}), we also measure Editing MOS (\textbf{EMOS}), which comprises a \textbf{Positive Rate} that is between 0 and 1 and measures whether or not the edited speech agrees with the editing, and the NMOS of the edited speech.}
\begin{tabular}{cc|ccccc}
\multicolumn{2}{c|}{\textbf{Model}}               & \textbf{Groundtruth}  & \textbf{Matcha-TTS}   & \textbf{Matcha-TTS-CFG} & \textbf{PPG2Speech}   & \textbf{PPG2Speech-CFG} \\ \hline
\multicolumn{2}{c|}{\textbf{NMOS}$\uparrow$}                & 4.27 $\pm$ 0.05 & 2.99 $\pm$ 0.08 & 2.96 $\pm$ 0.08   & 3.22 $\pm$ 0.08 & \textbf{3.51 $\pm$ 0.07}   \\
\multicolumn{2}{c|}{\textbf{SMOS}$\uparrow$}                & -            & \textbf{3.63 $\pm$ 0.09} & 3.39 $\pm$ 0.09   & 3.57 $\pm$ 0.09 & 3.54 $\pm$ 0.09   \\
\multicolumn{1}{c|}{\multirow{2}{*}{\textbf{EMOS}}} & \textbf{Positive Rate}$\uparrow$ & -            & -            & \textbf{0.87$\pm$ 0.02}    & -            & 0.81 $\pm$ 0.02   \\
\multicolumn{1}{c|}{}    & \textbf{NMOS}$\uparrow$          & -            & -            & 2.89 $\pm$ 0.05   & -            & \textbf{3.66 $\pm$ 0.06}  
\end{tabular}
\label{tab:sub_eval}
\end{table*}

In this editing experiment, we focus on single-phoneme replacement. We only edit single phoneme to simplify the evaluation process, and we omit deletions and insertions because Finnish is a phonetic language without silent phonemes, so it is unlikely for L2 speakers have phoneme deletions and insertions.

The results for editing experiments are shown in the \textbf{PAC} column in Table~\ref{tab:obj_eval}. The Matcha-TTS with CFG has a PAC of 0.804, while PPG2Speech with CFG has a PAC of 0.709. PPG2Speech achieves a relative $11.3\%$ reduction on the PAC.

\subsubsection{Subjective Evaluation}

The results for subjective evaluation are shown in Table~\ref{tab:sub_eval}. Our Mean Opinion Score (MOS) tests have 16 native Finnish participants for each test and 75 utterances for each system in each test. We use language proficiency questions that instruct the participants to rate certain questions with specific scores in Finnish text and audio prompts to filter out invalid submissions. Our MOS tests include naturalness MOS (\textbf{NMOS} row), speaker similarity MOS (\textbf{SMOS}), and a pronunciation editing MOS test (\textbf{EMOS} rows). For NMOS, the participants are asked to evaluate whether the speech is natural. For SMOS, the participants are presented with the natural reference utterances and synthetic speech with different content, and they are asked to rate how similar the speakers of these two speech samples are. The synthetic speech is synthesized using the same speaker embedding from the reference groundtruth. For both NMOS and SMOS, the scores are between 1-5 with an increment of 1. For EMOS, we provide the corresponding transcripts after the PPG editing, and the edited speech to the participants, while they are required to answer a yes-or-no question on whether the edited speech agrees with the edited transcripts, and rate the naturalness of the edited speech the same as NMOS. \textbf{Positive Rate} (the ratio of the \textit{yes} answer) is calculated as the subjective metric to the editing effect, which should be between 0 and 1. We use Student's t-test to obtain the $95\%$ confidence interval of each MOS, and use the Wilcoxon signed-rank test to determine whether the differences in MOS of any pair of systems are significant.

In Table~\ref{tab:sub_eval}, the Groundtruth has an NMOS of $4.27 \pm 0.05$, indicating the moderate dataset quality we have for this study. Among speech synthesis models, PPG2Speech-CFG achieves the highest NMOS ($3.51 \pm 0.07$). Moreover, comparing Matcha-TTS with Matcha-TTS-CFG using the Wilcoxon signed-rank test yields a p-value of $0.76$, meaning the difference on NMOS between Matcha-TTS and Matcha-TTS-CFG is insignificant, while the difference on NMOS between PPG2Speech and PPG2Speech-CFG is significant. This demonstrates that CFG can improve the naturalness in our experiments, and PPG2Speech-CFG can synthesize natural speech, and the slightly high CER shown in Table~\ref{tab:obj_eval} doesn't affect the listening experience. Matcha-TTS scores highest in SMOS ($3.63 \pm 0.09$), closely followed by PPG2Speech ($3.57 \pm 0.09$) and PPG2Speech-CFG ($3.54 \pm 0.09$). The Wilcoxon signed-rank test between the SMOS on PPG2Speech and the SMOS on PPG2Speech-CFG results in a p-value of $0.61$, suggesting the marginal advantage achieved by PPG2Speech over PPG2Speech-CFG is insignificant, while the Wilcoxon signed-rank test results between other pairs of SMOS suggest a significant difference. The SMOS is consistent with the SECS in Table~\ref{tab:obj_eval}, suggesting that pitch sequences can improve naturalness yet bring timbre leak to the model. To our surprise, Matcha-TTS-CFG shows the best Positive Rate ($0.87 \pm 0.02$) in the EMOS evaluation. We think this is caused by the undetermined nature of the PPG, which baffles the model from following the editing strictly. Meanwhile, PPG2Speech-CFG attains the highest NMOS within EMOS ($3.66 \pm 0.06$).

%% file: sections/Conclusion.tex
\section{Conclusions}

In this paper, we present PPG2Speech, a diffusion-based multispeaker PPG-to-Speech model, capable of phoneme-level pronunciation editing to approximate L2 language learners. We strengthen the original Matcha-TTS flow-matching decoder with CFG. We also proposed a task-specific objective evaluation metric, Phonetic Aligned Consistency (PAC), to measure the effectiveness of phoneme-level editing. 
The objective and subjective evaluations indicate that our PPG2Speech model can synthesize natural speech in both synthesis and editing tasks. It also follows the speaker identity in the speaker embedding, although timbre leaks caused slight degradation in the SECS and SMOS. For the editing task, both methods achieve more than $80\%$ positive rate in the test samples. Our PPG2Speech model performs better than the text-based editing in PAC, yet is inferior to the text-based editing in subjective evaluation, indicating that PAC might not be able to reflect the actual editing effect in some cases.
Our future work includes introducing foreign phonemes to the system and editing speech to mimic L2 speakers' pronunciation errors on phoneme duration and misplacement of open and closed vowels.

%% file: sections/acknowledge.tex
\section{Acknowledgements}

This work was a part of the Ministry of Education and Culture’s Doctoral Education Pilot in Finland, under Decision No. VN/3137/2024-OKM-6 (The Finnish Doctoral Program Network in Artificial Intelligence, AI-DOC). We acknowledge the computational resources provided by the Aalto Science-IT project.

We are grateful to Heini Kallio from Tampere University for her valuable discussion on the Finnish phonetics and common L2 mispronunciation, and to Antti Suni from University of Helsinki for sharing un the \textbf{Finsyn} dataset.